\documentclass[12pt]{article}
\usepackage[latin9]{inputenc}
\usepackage{amsmath}
\usepackage{amssymb}
\usepackage{graphicx}
\usepackage{esint}

\makeatletter

\newcommand\pubnumber{}
\newcommand\pubdate{}

\def\gll{\footnote{glin@cc.nctu.edu.tw}}
\def\yhl{\footnote{ravenchris.phys@gmail.com}}

\def\Title#1{\begin{center} {\Large #1 } \end{center}}
\def\Author#1{\begin{center}{ \sc #1} \end{center}}
\def\Address#1{\begin{center}{ \it #1} \end{center}}

\newcommand\pubblock{\rightline{\begin{tabular}{l} \pubnumber\\
         \pubdate  \end{tabular}}}
\newenvironment{Abstract}{\begin{quotation}  }{\end{quotation}}
\newenvironment{Presented}{\begin{quotation} \begin{center} 
             PRESENTED AT\end{center}\bigskip 
      \begin{center}\begin{large}}{\end{large}\end{center} \end{quotation}}
\def\Acknowledgements{\bigskip  \bigskip \begin{center} \begin{large}
             \bf ACKNOWLEDGEMENTS \end{large}\end{center}}

\def\beq{\begin{equation}}
\def\eeq#1{\label{#1}\end{equation}}
\def\eeqn{\end{equation}}

\def\beqa{\begin{eqnarray}}
\def\eeqa#1{\label{#1}\end{eqnarray}}
\def\eeqan{\end{eqnarray}}

\let\bar=\overbar

\def\Dslash{\not{\hbox{\kern-4pt $D$}}}
\def\dslash{\not{\hbox{\kern-2pt $\del$}}}

\def\msb{{\bar{\ssstyle M \kern -1pt S}}}

\makeatother

\begin{document}
\begin{titlepage} \pubblock

\vfill{}
\Title{Probing the coupling of heavy dark matter to nucleons by
detecting neutrino signature from the Earth core} \vfill{}
\Author{Guey-Lin Lin\gll and Yen-Hsun Lin\yhl} 

\Address{Institute of Physics, National Chiao Tung University\\
 Hsinchu 30010, Taiwan} \vfill{}
\begin{Abstract} We argue that the detection of neutrino signature
from the Earth core is an ideal approach for probing the coupling
of heavy dark matter ($m_{\chi}>10^{4}$ GeV) to nucleons. We first
note that direct searches for dark matter (DM) in such a mass range
do not provide stringent constraints. Furthermore the energies of
neutrinos arising from DM annihilations inside the Sun cannot exceed
a few TeV at the Sun surface due to the attenuation effect. Therefore
the sensitivity to the heavy DM coupling is lost. Finally, the detection
of neutrino signature from galactic halo can only probe DM annihilation
cross sections. After presenting the rationale of our studies, we
discuss the event rates in IceCube and KM3NeT arising from the neutrino
flux produced by annihilations of Earth-captured DM heavier than $10^{4}$
GeV. The IceCube and KM3NeT sensitivities to spin independent DM-proton
scattering cross section $\sigma_{\chi p}$ and isospin violation
effect in this mass range are presented. The implications of our results
are also discussed. \end{Abstract} \vfill{}
\begin{Presented} CosPA 2013\\
Symposium on Cosmology and Particle Astrophysics\\
Honolulu, Hawaii, November 12-15, 2013 \end{Presented} \vfill{}
\end{titlepage} \global\long\def\thefootnote{\fnsymbol{footnote}}
 \setcounter{footnote}{0}

\section{Introduction}

Evidences for the dark matter (DM) are provided by many astrophysical
observations, although the nature of DM is yet to be uncovered. DM
can be detected either directly or indirectly where the former observes
the nucleus recoil as DM interacts with the target nuclei in the detector
while the latter detects fi{}nal state particles resulting from DM
annihilations or decays. The current direct DM search limit on $\sigma_{\chi p}$
is up to 10 TeV mass only. Beyond 10 TeV mass, the indirect searches
such as IceCube and KM3NeT may use Earth as a target to probe $\sigma_{\chi p}$
.

The flux of DM induced neutrinos from galactic halo is only sensitive
to $\left\langle \sigma\upsilon\right\rangle $. Furthermore, the
energies of neutrinos from the Sun can not exceed a few TeVs due to
severe energy attenuations through the propagations. With the above
reasons, Earth is an ideal place to probe heavy DM couplings to nucleons.

In this work, we study both muon track events and cascade events induced
by neutrinos. We consider annihilation channels $\chi\chi\rightarrow\tau^{+}\tau^{-}$,
$W^{+}W^{-}$, and $\nu\bar{\nu}$ for signature neutrino productions.
Recent studies \cite{Kurylov:2003ra,Feng:2011vu} also suggested that
DM-nucleon interactions do not necessarily respect isospin symmetry.
Therefore isospin violation effect is also taken into consideration
in our analysis.

\section{Neutrino signals from DM and atmospheric background}

\subsection{DM capture and annihilation rates in the Earth core}

The neutrino differential flux $\Phi_{\nu_{i}}$ from $\chi\chi\rightarrow f\bar{f}$
can be expressed as
\begin{equation}
\frac{d\Phi_{\nu_{i}}}{dE_{\nu_{i}}}=P_{\nu_{j\rightarrow i}}(R_{\oplus},E_{\nu})\frac{\Gamma_{A}}{4\pi R_{\oplus}^{2}}\sum_{f}B_{f}\left(\frac{dN_{\nu_{j}}}{dE_{\nu_{j}}}\right)_{f}\label{eq:neutrino_flux}
\end{equation}
where $R_{\oplus}$ is the Earth radius, $P_{\nu_{j\rightarrow i}}$
is the neutrino oscillation probability from flavor $j$ to $i$ after
propagating from the source to the detector, $B_{f}$ is the branching
ratio corresponding to the channel $\chi\chi\rightarrow f\bar{f}$
, $dN_{\nu}/dE_{\nu}$ is the neutrino spectrum, and $\Gamma_{A}$
is the DM annihilation rate in the Earth.

The annihilation rate, $\Gamma_{A}$, can be obtained by solving the
DM evolution equation in the Earth core \cite{Olive:1986kw,Srednicki:1986vj}
\begin{equation}
\dot{N}=\Gamma_{C}N-C_{A}N^{2}-C_{E}N\label{eq:evolution_eq}
\end{equation}
where $N$ is the DM number density in the Earth core, $\Gamma_{C}$
is the capture rate, and $C_{E}$ is the evaporation rate. The evaporation
rate is only relevant when $m_{\chi}\lesssim5\textrm{ GeV}$ \cite{Gould:1987ir}
and can be ignored in this work. Solving Eq.~\eqref{eq:evolution_eq}
thus gives
\begin{equation}
\Gamma_{A}=\frac{C_{A}}{2}N(t)^{2}=\frac{\Gamma_{C}}{2}\tanh^{2}\left(\frac{t}{\tau_{\oplus}}\right)\label{eq:annihilation_rate}
\end{equation}
where $t$ is the lifetime of the solar system and $\tau_{\oplus}$
is the time scale when the DM capture and annihilation in the Earth
core reaches the equilibrium state. The capture rate, $\Gamma_{C}$,
is proportional to
\begin{equation}
\Gamma_{C}\propto\left(\frac{\rho_{0}}{0.3\textrm{ GeV cm}^{-3}}\right)\left(\frac{270\textrm{ km s}^{-1}}{\bar{\upsilon}}\right)\left(\frac{\textrm{GeV}}{m_{\chi}}\right)\left(\frac{\sigma_{\chi p}}{\textrm{pb}}\right)\sum_{\mathcal{A}}F_{\mathcal{A}}^{*}(m_{\chi})\label{eq:capture_rate}
\end{equation}
where $\rho_{0}$ is the local DM density, $\bar{\upsilon}$ is the
DM velocity dispersion, $\sigma_{\chi p}$ is the DM-nucleon cross
sections, and $F_{\mathcal{A}}^{*}(m_{\chi})$ is the product of various
factors for element $\mathcal{A}$ including the mass fraction, chemical
element distribution, kinematic suppression, form-factor and reduced
mass.

\subsubsection{The effect of isospin violation}

Given an element with atom number $A$, atomic number $Z$ and the
reduced mass of the element and DM particle $\mu_{A}=m_{\chi}m_{A}/(m_{\chi}+m_{A})$.
By assuming $m_{p}\approx m_{n}$, the usual DM-nucleus cross section
is written as \cite{Jungman:1995df},
\begin{equation}
\sigma_{\chi A}=\frac{4\mu_{A}^{2}}{\pi}[Zf_{p}+(A-Z)f_{n}]^{2}=A^{2}\left(\frac{m_{\chi}+m_{p}}{m_{\chi}+m_{A}}\right)^{2}\left[Z+(A-Z)\frac{f_{n}}{f_{p}}\right]^{2}\sigma_{\chi p}.
\end{equation}
where $\sigma_{\chi p}$ is the DM-proton scattering cross section.
If the effective couplings of DM to protons, $f_{p}$, and neutrons,
$f_{n}$, are not identical, the capture rate, Eq.~\eqref{eq:capture_rate},
becomes
\begin{equation}
\Gamma_{C}^{\textrm{IV}}\propto\xi(\rho_{0},\bar{\upsilon},m_{\chi})\left(\frac{\sigma_{\chi p}^{\textrm{IV}}}{\textrm{pb}}\right)\sum_{\mathcal{A}}F_{\mathcal{A}}^{*}(m_{\chi})A^{2}\left(\frac{m_{\chi}+m_{p}}{m_{\chi}+m_{A}}\right)^{2}\left[Z+(A-Z)\frac{f_{n}}{f_{p}}\right]^{2}\label{eq:IV_capture_rate}
\end{equation}
where $\xi(\rho_{0},\bar{\upsilon},m_{\chi})$ is the first three
terms in Eq.~\eqref{eq:capture_rate}. The superscript IV stands
for isospin violation. It is important to note that the $\sigma_{\chi p}^{\textrm{IV}}$
here is the DM-proton cross section derived from isospin violation
condition and not identical to the $\sigma_{\chi p}$ in Eq.~\eqref{eq:capture_rate}
in general.

\subsubsection{Neutrino signal and atmospheric background event rates}

The neutrino event rate in the detector from the Earth DM is given
by
\begin{equation}
N_{\nu}=\int_{E_{\textrm{th}}}^{m_{\chi}}\frac{d\Phi_{\nu}}{dE_{\nu}}A_{\nu}(E_{\nu})dE_{\nu}d\Omega\label{eq:nu_event}
\end{equation}
where $E_{\textrm{th}}$ is the detector threshold energy, $d\Phi_{\nu}/dE_{\nu}$
is the neutrino flux from DM annihilations, $A_{\nu}$ is the detector
effective area \cite{Aartsen:2013uuv,Aartsen:2013jdh,Katz:2011zza},
and $\Omega$ is the solid radian. The atmospheric background event
rate has a similar expression,
\begin{equation}
N_{\textrm{atm}}=\int_{E_{\textrm{th}}}^{E_{\textrm{max}}}\frac{d\Phi_{\nu}^{\textrm{atm}}}{dE_{\nu}}A_{\nu}(E_{\nu})dE_{\nu}d\Omega.\label{eq:atm_event}
\end{equation}
We set $E_{\textrm{max}}=m_{\chi}$ in Eq.~\eqref{eq:atm_event}
to compare with the DM signal.

\section{Results\label{sec:Results}}

We present the sensitivity as a $2\sigma$ detection significance
in 5 years, calculated with the convention,
\begin{equation}
\frac{s}{\sqrt{s+b}}=2.0\label{eq:convention_eq}
\end{equation}
where $s$ is the DM signal, $b$ the atmospheric background, and
$2.0$ referring to the $2\sigma$ detection significance. The atmospheric
$\nu_{\tau}$ flux is extremely small and can be ignored in our analysis.
Thus we take $\nu_{e}$ and $\nu_{\mu}$ as our major background sources.
The detector threshold energy $E_{\textrm{th}}$ in Eq.~\eqref{eq:nu_event}
and \eqref{eq:atm_event} is set to be $10^{4}$ GeV in order to suppress
the incoming background. In our analysis, we present two isospin scenarios
for the constraints on $\left\langle \sigma\upsilon\right\rangle $
and $\sigma_{\chi p}$. One is $f_{n}/f_{p}=1$, the isospin symmetry
case, and the other is $f_{n}/f_{p}=-0.7$, the isospin violation
one. 

To constrain DM-annihilation cross section $\left\langle \sigma\upsilon\right\rangle $,
we make use of the $\sigma_{\chi p}$ from the extrapolation of the
LUX bound \cite{Akerib:2013tjd} to $m_{\chi}>10$ TeV.

\subsection{IceCube sensitivities}

\begin{figure}[t]
\begin{centering}
\includegraphics[width=0.49\textwidth]{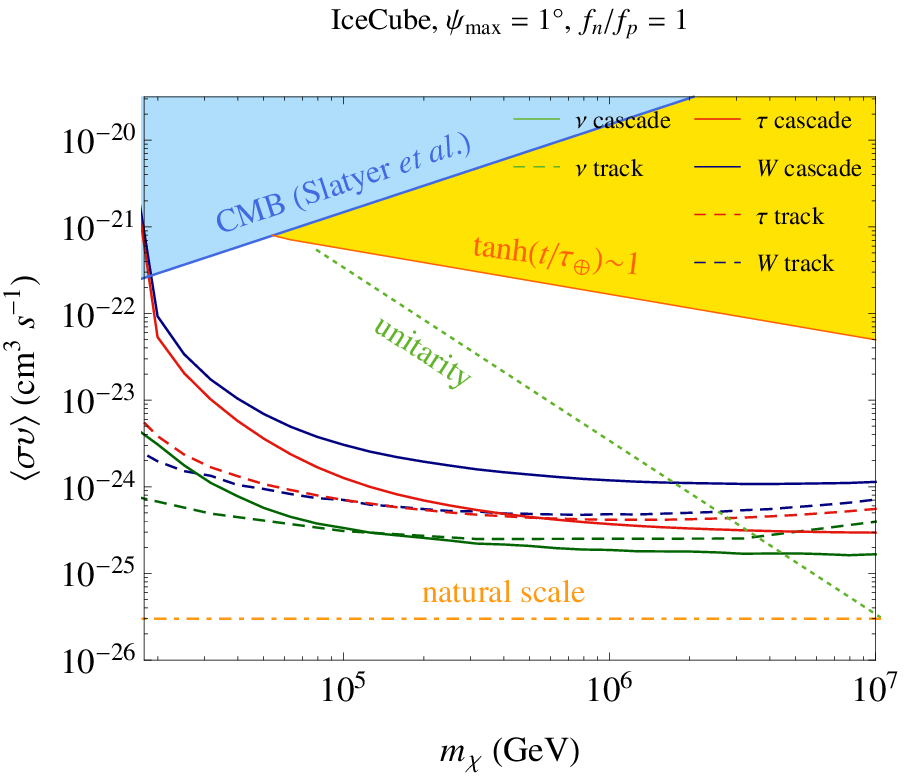}\includegraphics[width=0.49\textwidth]{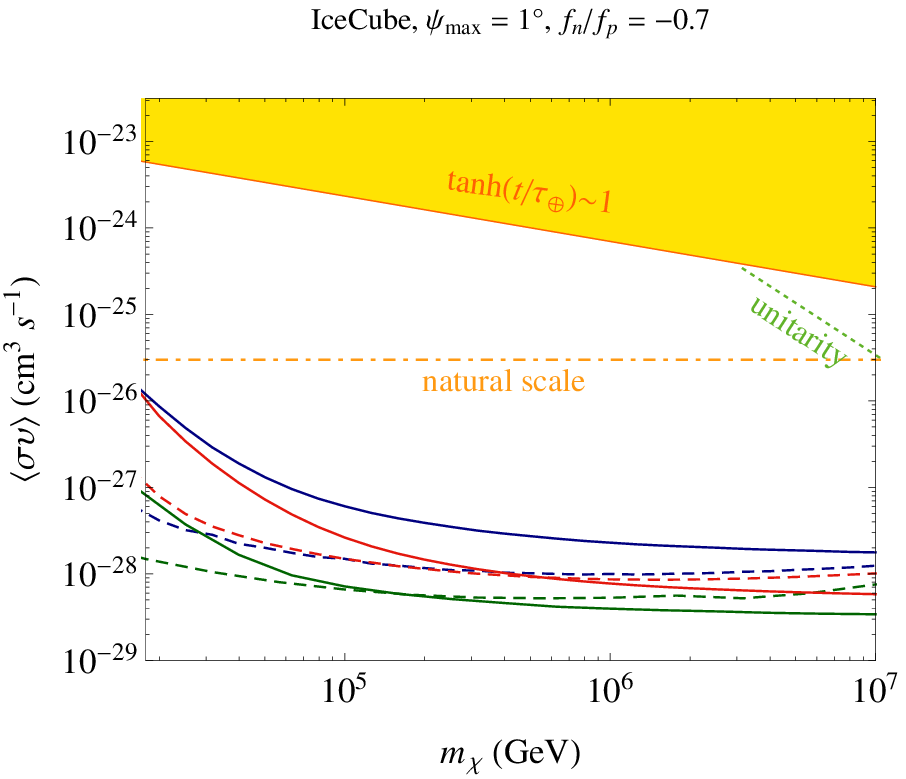}
\par\end{centering}

\caption{\label{fig:IceCube_SV}The IceCube 5-year sensitivity at $2\sigma$
to $\left\langle \sigma\upsilon\right\rangle $ for $\chi\chi\rightarrow\tau^{+}\tau^{-}$,
$W^{+}W^{-}$, and $\nu\bar{\nu}$ annihilation channels with track
and cascade events, respectively. The isospin symmetry case, $f_{n}/f_{p}=1$,
is presented on the left panel, and the isospin violation case, $f_{n}/f_{p}=-0.7$,
is presented on the right panel. The yellow-shaded region is the parameter
space for the equilibrium state and the blue-shade region is the constraint
from CMB \cite{Slatyer:2009yq}. }
\end{figure}
\begin{figure}[t]
\begin{centering}
\includegraphics[width=0.49\textwidth]{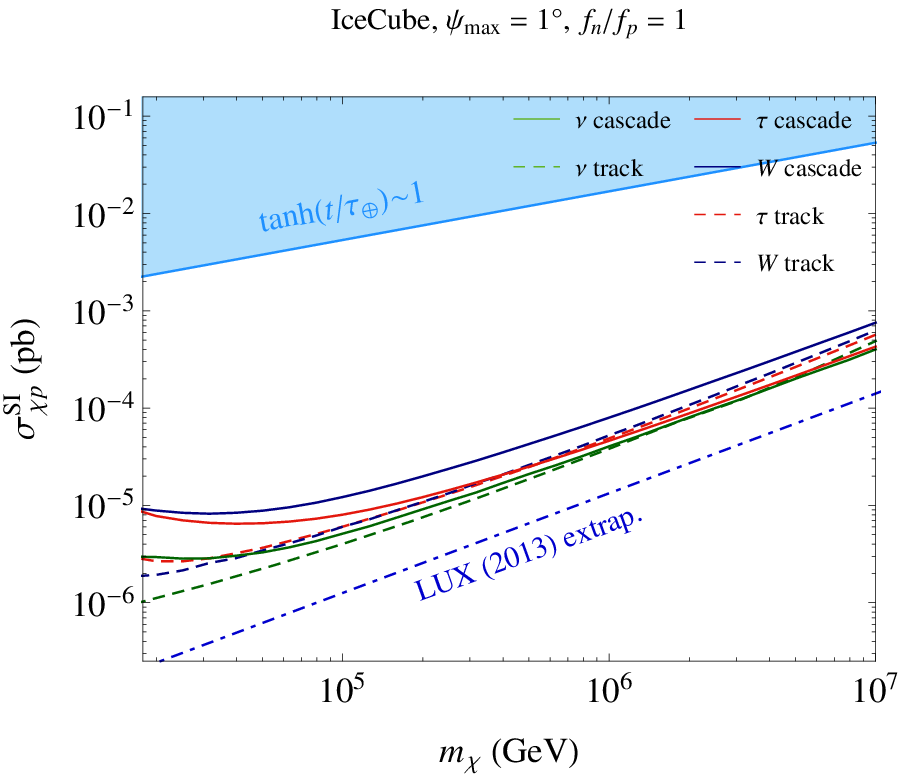}\includegraphics[width=0.49\textwidth]{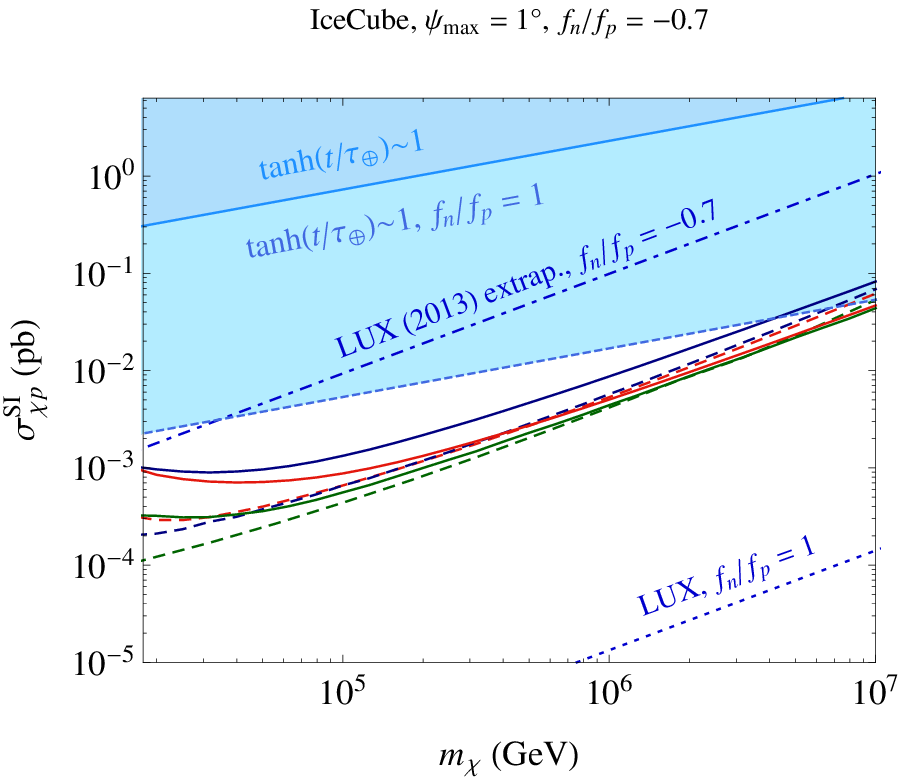}
\par\end{centering}

\caption{\label{fig:IceCube_SI}The IceCube $2\sigma$ sensitivities in 5 years
to $\sigma_{\chi p}^{\textrm{SI}}$ for $\chi\chi\rightarrow\tau^{+}\tau^{-}$,
$W^{+}W^{-}$, and $\nu\bar{\nu}$ annihilation channels with both
track and cascade events. The isospin symmetry case, $f_{n}/f_{p}=1$,
is presented on the left panel, and the isospin violation case, $f_{n}/f_{p}=-0.7$,
is presented on the right panel. The blue-shaded region is the parameter
space for the equilibrium state and the light-blue-shaded region on
the right panel refers to the equilibrium-state parameter space for
the isospin symmetry case as a comparison. An extrapolation of current
LUX limit has been shown on the figures.}
\end{figure}

In Fig.~\ref{fig:IceCube_SV} we present the IceCube sensitivities
to $\left\langle \sigma\upsilon\right\rangle $ of $\chi\chi\rightarrow\tau^{+}\tau^{-}$,
$W^{+}W^{-}$, and $\nu\bar{\nu}$ annihilation channels in the Earth
core with both track and cascade events. In $f_{n}/f_{p}=1$, the
IceCube sensitivities to track events from $\chi\chi\rightarrow\tau^{+}\tau^{-}$
and $W^{+}W^{-}$ annihilation channels are comparable while one expects
to obtain the most stringent constraint on the annihilation cross
section by analyzing track and cascade events from $\chi\chi\rightarrow\nu\bar{\nu}$.

However, the isospin violation scenario, $f_{n}/f_{p}=-0.7$, will
weaken the LUX bound by 4 orders of magnitude, i.e., the LUX upper
bound on $\sigma_{\chi p}$ is raised by 4 orders of magnitude. With
a 4-order larger $\sigma_{\chi p}$, the capture rate given by Eq.~\eqref{eq:IV_capture_rate}
is enhanced by 2 orders of magnitude since the suppression factor
due to the isospin violation is around $10^{-2}$ for chemical elements
in the Earth core. With the capture rate enhanced by 2 orders of magnitude,
the IceCube sensitivities to $\left\langle \sigma\upsilon\right\rangle $
of various annihilation channels can be improved by about 4 orders
of magnitude. Therefore, the sensitivities could reach below the natural
scale $\left\langle \sigma\upsilon\right\rangle =3\times10^{-26}\textrm{ cm}^{2}\textrm{ s}^{-1}$.

Fig.~\ref{fig:IceCube_SI} shows the IceCube sensitivities to spin-independent
cross section $\sigma_{\chi p}^{\textrm{SI}}$ by analyzing track
and cascade events from $\chi\chi\rightarrow\tau^{+}\tau^{-}$, $W^{+}W^{-}$,
and $\nu\bar{\nu}$ annihilation channels in the Earth core. The threshold
energy $E_{\textrm{th}}$ is the same as before and we take $\left\langle \sigma\upsilon\right\rangle =3\times10^{-26}\textrm{ cm}^{2}\textrm{ s}^{-1}$
as our input. Precisely speaking, the sensitivity to $\chi\chi\rightarrow\nu\bar{\nu}$
channel is the highest. However, the sensitivities to different channels
can be taken as comparable since the differences between them are
not significant.

When isospin is a good symmetry, the IceCube sensitivities are no
better than constraints from the LUX extrapolation. However, with
$f_{n}/f_{p}=-0.7$, the capture rate in Eq.~\eqref{eq:IV_capture_rate}
is reduced to 1\% of the isospin symmetric value. Therefore one requires
100 times larger $\sigma_{\chi p}^{\textrm{SI}}$ to reach the same
detection significance. However, the ratio $f_{n}/f_{p}=-0.7$ makes
a more dramatic impact to the DM direct search using xenon as the
target. The DM scattering cross section with xenon is reduced by 4
orders of magnitude. Hence the indirect search by IceCube could provide
better constraint on $\sigma_{\chi p}^{\textrm{SI}}$ than the direct
search in such a case.

\subsection{KM3NeT sensitivities}

\begin{figure}[t]
\begin{centering}
\includegraphics[width=0.49\textwidth]{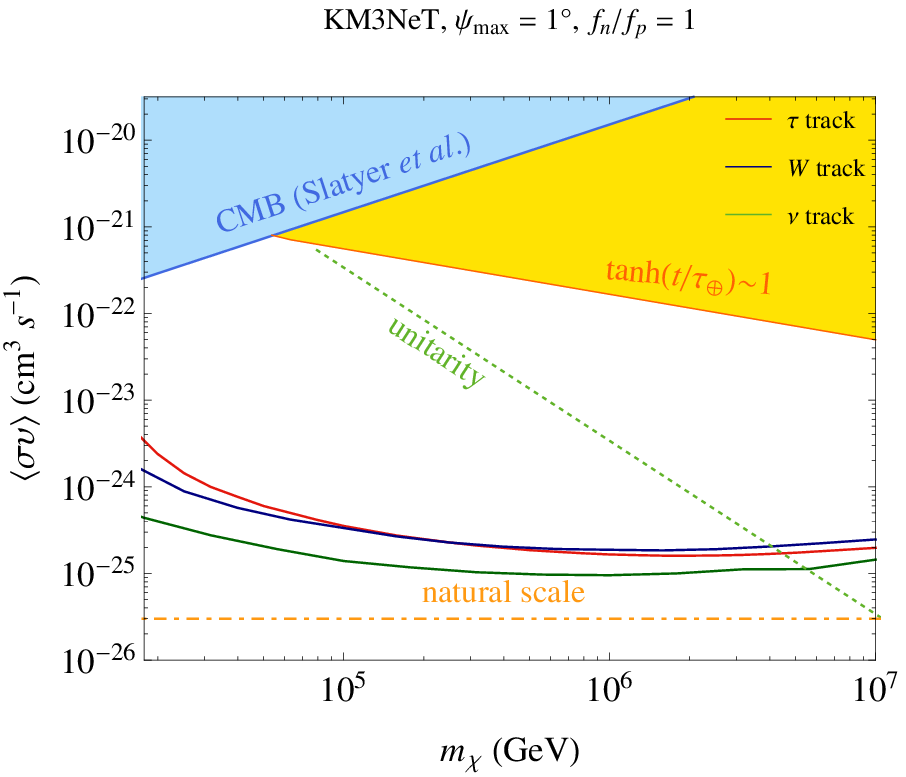}\includegraphics[width=0.49\textwidth]{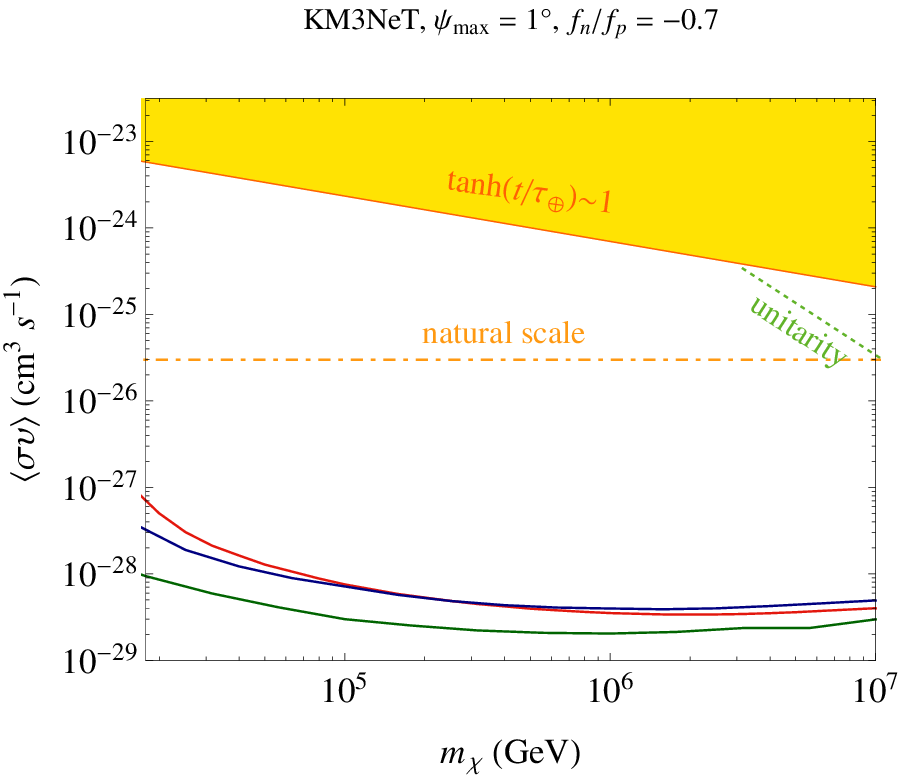}
\par\end{centering}

\caption{\label{fig:KM3SV}The KM3NeT $2\sigma$ sensitivities in 5 years to
$\left\langle \sigma\upsilon\right\rangle $ for $\chi\chi\rightarrow\tau^{+}\tau^{-}$,
$W^{+}W^{-}$, and $\nu\bar{\nu}$ annihilation channels with track
events only. The isospin symmetry case, $f_{n}/f_{p}=1$, is presented
on the left panel, and the isospin violation case, $f_{n}/f_{p}=-0.7$,
is presented on the right panel. The yellow-shaded region is the parameter
space for the equilibrium state and the blue-shaded region is the
constraint from CMB \cite{Slatyer:2009yq}.}
\end{figure}
\begin{figure}[t]
\begin{centering}
\includegraphics[width=0.49\textwidth]{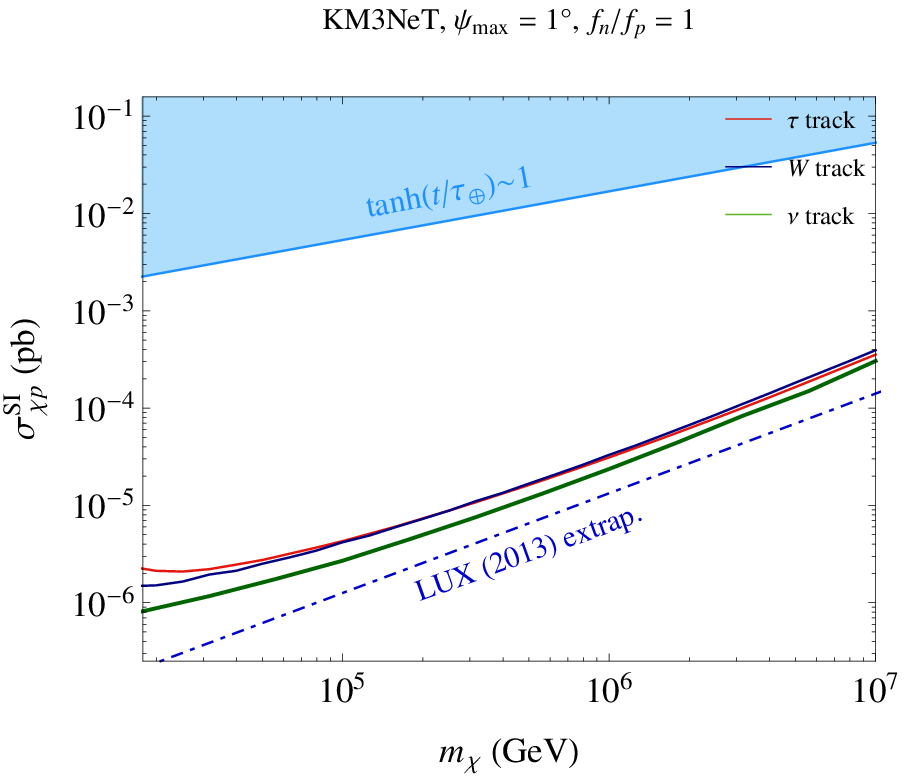}\includegraphics[width=0.49\textwidth]{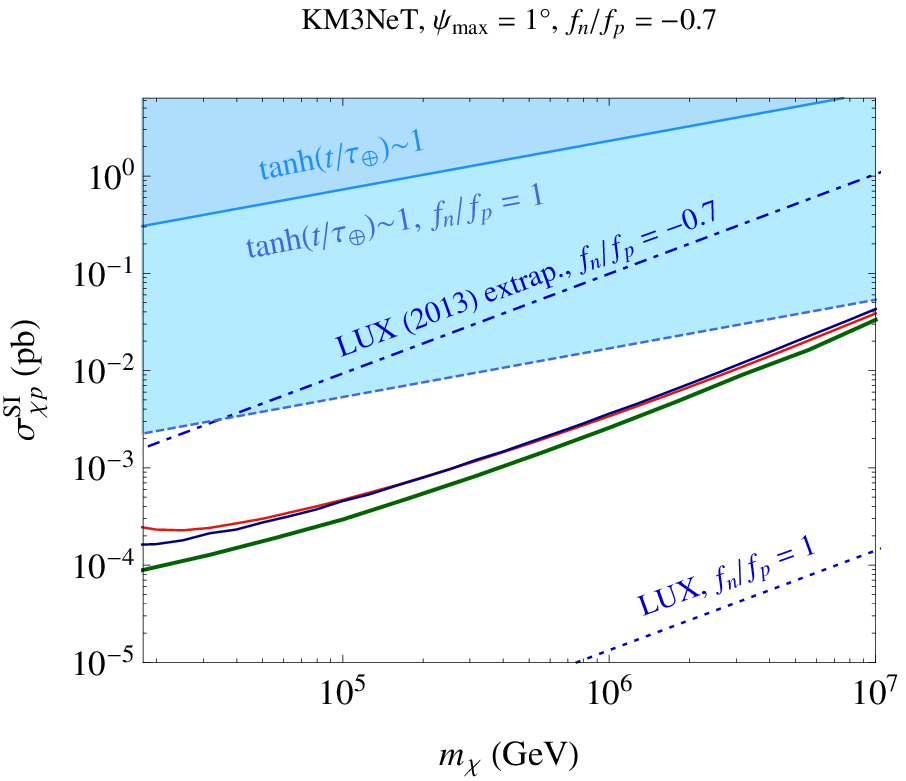}
\par\end{centering}

\caption{\label{fig:KM3SI}The KM3NeT $2\sigma$ sensitivities in 5 years to
$\sigma_{\chi p}^{\textrm{SI}}$ for $\chi\chi\rightarrow\tau^{+}\tau^{-}$,
$W^{+}W^{-}$, and $\nu\bar{\nu}$ annihilation channels for track
events only. The isospin symmetry case, $f_{n}/f_{p}=1$, is presented
on the left panel, and the isospin violation case, $f_{n}/f_{p}=-0.7$,
is presented on the right panel. The blue-shaded region is the parameter
space for the equilibrium state and the light-blue-shaded region on
the right panel refers to the equilibrium-state parameter space in
the isospin symmetry case.}
\end{figure}

Besides IceCube, the neutrino telescope KM3NeT located in the northern-hemisphere
shall also reach to a promising sensitivity in the near future. Therefore
it is worthwhile to comment on the performance of KM3NeT. Since KM3NeT
only publishes $\nu_{\mu}$ charge-current effective area in the present
stage, we shall only analyze track events.

The results are shown in Fig.~\ref{fig:KM3SV} and \ref{fig:KM3SI}
with parameters chosen to be the same as those for computing the IceCube
sensitivities. The KM3NeT sensitivities are almost 1 order of magnitude
better than the IceCube ones due to its $\nu_{\mu}$ C.C. effective
area is about one order of magnitude larger than IceCube's.

\section{Summary}

In this work we have presented the IceCube and KM3NeT sensitivities
to DM spin-independent cross section $\sigma_{\chi p}$ and annihilation
cross section $\left\langle \sigma\upsilon\right\rangle $ by detecting
DM induced signature from the Earth's core. The direct DM search only
probes $\sigma_{\chi p}$ with the sensitivity dropping quickly with
DM mass for $m_{\chi}>10^{4}$ GeV. However, the indirect search using
the large underground neutrino telescopes such IceCube or KM3NeT could
probe $\sigma_{\chi p}$ in such a mass range. Besides, the indirect
search can also probe $\left\langle \sigma\upsilon\right\rangle $.

We have also shown that, like the direct search, the indirect search
is affected by the isospin violation. The implications of isospin
violation to IceCube and KM3NeT observations have been presented in
Sec.~\ref{sec:Results}. Taking isospin violation effect into account,
the sensitivities of the above neutrino telescopes to $\sigma_{\chi p}$
and $\left\langle \sigma\upsilon\right\rangle $ for different channels
could be better than the direct search limit and the natural scale,
respectively for a certain range of $f_{n}/f_{p}$.

\Acknowledgements

This work is supported by National Science Council of Taiwan under
Grant No.~102-2112-M-009-017. 



\begin{thebibliography}{99}

\bibitem{Kurylov:2003ra}
A.~Kurylov and M.~Kamionkowski,
Phys.\ Rev.\ D {\bf 69}, 063503 (2004). 

\bibitem{Feng:2011vu} 
J.~L.~Feng, J.~Kumar, D.~Marfatia and D.~Sanford,
Phys.\ Lett.\ B {\bf 703}, 124 (2011). 

\bibitem{Olive:1986kw}
K.~A.~Olive, M.~Srednicki and J.~Silk,
UMN-TH-584/86.

\bibitem{Srednicki:1986vj}
M.~Srednicki, K.~A.~Olive and J.~Silk,
Nucl.\ Phys.\ B {\bf 279}, 804 (1987).

\bibitem{Gould:1987ir} 
A.~Gould,
Astrophys.\ J.\  {\bf 321}, 571 (1987).

\bibitem{Jungman:1995df}
G.~Jungman, M.~Kamionkowski and K.~Griest,
Phys.\ Rept.\ {\bf 267}, 195 (1996). 

\bibitem{Aartsen:2013uuv}
M.~G.~Aartsen {\it et al.} [IceCube Collaboration],
arXiv:1307.6669 [astro-ph.HE].

\bibitem{Aartsen:2013jdh}
M.~G.~Aartsen {\it et al.}  [IceCube Collaboration],
Science {\bf 342}, No. 6161, 1242856 (2013). 

\bibitem{Katz:2011zza}
U.~F.~Katz [KM3NeT Collaboration],
Nucl.\ Instrum.\ Meth.\ A {\bf 626-627}, S57 (2011).

\bibitem{Akerib:2013tjd}
D.~S.~Akerib {\it et al.}  [LUX Collaboration],
Phys.\ Rev.\ Lett.\ {\bf 112}, 091303 (2014). 

\bibitem{Slatyer:2009yq}
T.~R.~Slatyer, N.~Padmanabhan and D.~P.~Finkbeiner,
Phys.\ Rev.\ D {\bf 80}, 043526 (2009). 

\end{thebibliography}
\end{document}